# Evaluation of Game Design Framework Using a Gamified Browser-Based Application

*Abdulrahman Hassan Alhazmi, The Department of Computer Science and IT at La Trobe University, Australia and Jazan University, Saudi Arabia*
*Nalin Asanka Gamagedara Arachchilage, The School of Computer Science, The University of Auckland, New Zealand and La Trobe University, Australia*

## Abstract

Privacy Policy under GDPR law helps users understand how software developers handle their personal data. GDPR privacy education must be considered a vital aspect of combating privacy threats. In this paper, we present the design and development of a gamified browser-based application aimed at motivating software developers to enhance their secure coding behavior. To evaluate the proposed game design framework through the developed framework, a think-aloud study was carried out along with pre-and post-test. There was an improvement in the software developers' secure coding behaviors through playing the game which had GDPR privacy laws incorporated to enhance their knowledge of privacy.

## 1. Introduction

The extensive use of software applications continues to pose a threat to user privacy when interacting with software systems [1][2][22]. Although privacy policies like Privacy by Design (PbD) give software developers explicit guidance on incorporating privacy into software designs, these practices have not yet become widespread among software developers [13]. There are other practices other than lack of knowledge that hinder developers from implementing privacy, such as lack of resources, regulations, and incentives [36].

Numerous studies have been conducted on software users' views and concerns about privacy [2][3][4][5][6]. Popular models that represent users' privacy decision-making are the result of user-centric research [2][7][8][4]. Webmasters' [9] and IT administrators' [10] perspectives of surveillance were the subject of several studies. As requirements are established, and a suitable solution for achieving them is built, which in this case affects how the system collects and handles personal data, it is still unclear how privacy fits into the system design process [11]. In particular, our knowledge of how developers approach and understand informational privacy is extremely limited, as is the knowledge of policymakers who support PbD.

In order to safeguard the protection of EU citizens' personal data, the General Data Protection Regulation (GDPR) [14], which has been in effect since 2016 and became mandatory in May 2018, establishes a novel range of legally binding data protection laws, data subjects' rights, and obligations. As is often remarked, "[software] code is law," meaning that the support of technological features controls what we can do just as much as the legal framework. Such a "privacy-by-policy" [15] approach, though, places the burden of adhering to regulations in the hands of legal staff, leaving engineers ill-equipped to deal with related concepts and without the necessary resources to incorporate those regulations into the software products they design.

Numerous approaches [32] have been extensively investigated and developed in the purely technological field to produce Privacy-Enhancing Technologies (PETs) with varied degrees of maturity which address privacy issues. The approaches were extensively researched since they will provide satisfactory protection of privacy [32]. The majority of software engineers are still unaware of privacy-enhancing technologies [33] since there is limited access and minimal education and training; however, as a result of the separation between systematic engineering and development and PETs, which prevents software engineers from understanding or being aware of the right applicability of such solutions.

Software tools for privacy management are designed to make it easier for non-experts to comply with new requirements like GDPR [26]. Educating developers about GDPR law through gamification enable interactive learning and increases attentiveness [34] will, in turn, improve their motivation to protect private user data and improve their secure coding behavior as they write software code. Developers' motivation will be high since they will come up with different ways of enabling the privacy of software applications.

The proposed game design approach (shown in Figure 1) aims to increase software developers' motivational coding habits toward privacy. The GDPR principles [14] and a previously developed game design framework [16] are all combined in this model to teach software developers how to incorporate privacy into the software systems they are developing.

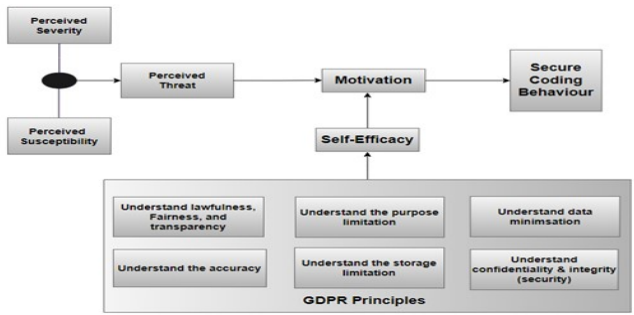

Figure 1: The Proposed Serious Game Design Framework

## 2. Related Work

Research has proven that technology alone cannot address all critical privacy issues. Literature provides numerous definitions and viewpoints on the very broad and varied concept of privacy [17]. Historically, privacy has been viewed differently in terms of media, territory, communication, and bodily privacy. The exorbitant number of data breaches—4,700 of which have been made public in the US alone since 2005—shows the effects quite evidently [19]. A number of studies examined the development and implementation of privacy-by-design systems, especially from the perspective of requirements engineering, where privacy is seen as a problem of compliance or the stringent and formally validated observance of rules and by-laws [20].

The General Data Protection Policy (Regulation (EU) 2016/679), passed on April 14, 2016, and went into effect on May 25, 2018, [14] is the European Union's approach to online privacy. It is the most comprehensive privacy regulation to date. Data subjects, data controllers, data processors, and third parties are the four categories of entities defined by the GDPR. Information systems users from where data is collected are considered the data subjects. The service provider (such as a website or mobile application) with a legitimate interest in acquiring and processing user data is normally the data controller. A data controller may hire a processor to handle data processing on its behalf. The data controller may also permit a third party (such as an analytics company) to process some of the user's data [21]. Orwell, a game on data privacy, enables a developer to play as a government worker to protect their country through surveillance of online activities and gather information [37]. Orwell software game does not consider all our GDPR principles, making our application more suitable.

Previous studies were performed to determine what obstacles developers face when adding privacy protections to software programs [2][3][4][5][6]. Lack of awareness, training, and education on privacy laws, especially GDPR principles, among developers has resulted in less privacy for end-users [35]. In this paper, we present the design and development of a gamified browser-based application as an educational intervention to motivate software developers to enhance their secure coding behavior. Alhazmi and Arachchilage [30] concluded from our research that one of the barriers to GDPR onboarding is developers' lack of familiarity with GDPR principles [29], which directly impacts the developers' coding behavior. A game design framework introduced by [16] assessed users' phishing threat avoidance behavior by using a game-based anti-phishing approach; however, no work has been done in order to develop a framework through an empirical investigation that motivates developers to embed privacy into software and hence improve their secure coding behavior.

## 3. Game Design Issues

The primary goal of the proposed browser-based gamified application is to educate programmers on how to incorporate GDPR privacy laws into their software applications.

By employing game-based GDPR education, a game design framework developed by [22] and validated by [28] analyzed developers' Secure Coding Behavior. Several components of their structure were included in our design.

According to the game design framework presented in Figure 1, developers' secure coding behavior is influenced by their motivation, which is, in turn, influenced by a perceived threat. Perceived severity, susceptibility, and their combination (interaction effect) all affect the perceived threat. While self-efficacy also affects developers' motivation. While the game design framework highlights the issues that the game design must solve, it should also guide how to organize this data and present it within the context of a game. In order to do this, we set out to create a threat perception that would encourage developers to apply GDPR.

### 3.1. What to Teach

Developers use two Six guiding principles of GDPR to implement privacy. The framework operates by first identifying how developers (game players) view potential dangers to data privacy. The game design framework educates the following GDPR principles to make sure that developers are aware of the privacy threats:

**Lawfulness, fairness, and transparency**: The GDPR mandates that data subjects be informed about how their personal information is handled. The developer should also grasp the consents obtained by a user before processing data. The developer also needs to be open and honest with the user about how data will be utilized [22]. **Purpose limitation**: Developers must adhere to the principle of compatibility, which forbids them from processing user data for purposes unrelated to those for which it was originally intended. In addition, developers should uphold their integrity by not exploiting client data in incompatible ways without their consent [22]. **Data minimization**: Before gathering data, a developer should first grasp its importance; if it is not nec-essary to finish a particular operation, it should not be gathered. Second, a developer should be aware that to perform a task, s/he should process as little information as feasible [22]. **Accuracy**: A developer should be aware of the authenticity of user data, for instance, through verification of the data collected [22]. **Storage limitation**: A player should be aware of the data retention policy, which states that data shouldn't be kept around if it's not being used [22]. **Integrity and confidentiality (security)**: To achieve integrity, data must retain its accuracy after transmission. Data protection against unauthorized access is necessary for confidentiality and can be achieved through encryption, access control, proper training, awareness and being pseudonymous [22].

### 3.2. Story and Mechanism

In this game design scenario, the software developer is the player. In the beginning, a gameplay environment is provided to the game player illustrating specific scenarios of data breaches caused by software systems that were probably coded badly. The game should be designed in such a way that players (developers) sense a threat to data privacy that could affect the user data and come up with a software solution that protects data privacy if they learn this through the game. As players discover how software developers decide whether to apply GDPR principles to software systems, the game should also increase the developers' self-efficacy. Understanding how the GDPR improves developers' capability to make decisions on how to protect these data items (i.e., when collecting, storing, and sharing) during the development of software systems that protect user privacy.

When a threat is encountered, the developer will have to address it at all levels of applying safety, taking into account the cost, effectiveness, and adherence to GDPR standards. In order to succeed in the game, the player must follow the rules and perform the appropriate actions to deal with any threats that may arise while taking into account game elements (time and money). This entire evaluation will encourage the developer (gamer) to develop a software application that preserves privacy.

### 3.3 The Gamified Application

To explore the possibility of using a game to provide knowledge of six GDPR principles to the developers, a working model of a gamified browser-based application was developed. The game has six levels and videos from "start" to "end," as shown in Figure 2. In each game video, an objective scenario regarding the six GDPR principles is given. Each video of the game contains a conversation between a software developer, cyber security consultant, doctor, nurse, and patient on the principles and how they are applied and implemented in developing a Health Information System that preserves end-user privacy. Also, at each level of the game, an objective type of question regarding the six GDPR principles is asked. Each level of the game contains three questions. A total of 60 seconds is allotted to game players to answer as many questions as they can. A bonus of 5 seconds is earned with each correct answer. Players must answer one question correctly to earn a star, two questions to earn two stars, and three questions to earn all three stars to get a perfect score and to move on to the next level.

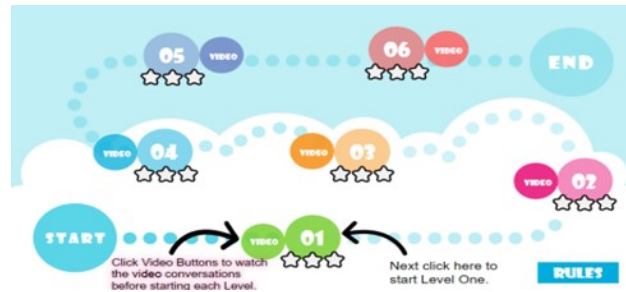

Figure 2: The GDPR gamified browser-based application

## 4. Methodology and Research Design

A think-aloud experiment was done along with a pre-test, play game, and post-test to evaluate a mobile game prototype for developers to enhance the implementation of GDPR principles in their coding behavior to protect end-user privacy. First, we asked developers to take a pre-test with six sections to get their understanding of GDPR principles. Each section has three questions regarding GDPR principles and videos. While playing the game, we asked questions (Exit Questions) to see if developers transfer the concept of GDPR principles into the real programming world. Then, to determine the level of subjective satisfaction with the gamified application, our study used a usability study of the application. Finally, we asked developers to take a post-test to determine and measure their understanding of GDPR principles after getting a full understanding of GDPR principles and their threats. The think-aloud experiment allowed the developers to come up with views touching on privacy and analyze those views based on how they found the game.

### 4.1. Data Collection

Both qualitative and quantitative data collection techniques were used in this study. To achieve this goal, we used the System Usability Scale (SUS), a tool for measuring users' subjective satisfaction with the application's usability [23]. Sample sizes of at least 12–14 participants are recommended by [24] and [25] to obtain reasonably reliable results under the study conditions. The SUS employs a five-point Likert scale with strongly agree and strongly disagree anchors.

In addition to the pre-and post-test, a think-aloud study protocol was used to gather information on how the user's interaction with the gamified application affected the elements of the gamified application. In the second stage of the think-aloud study, developers were asked exit questions during the gameplay activity. This provided us with an insight into the developers coding behavior before and after playing the game.

### 4.2. Main Study

#### 4.2.1 Participants

A total number of 20 participants took part in the think-aloud study, which included pre-and post-test, to see how well they understood and knew about GDPR law using the gamified application.

To participate in this study, participants had to meet the eligibility criteria of being software developers and designers. Developers were contacted through LinkedIn, and after being identified, they were sent an email invitation. The participant information consent form was emailed to participants, and the appropriate box was checked by those who were willing to participate. After providing their permission to take part in the study, participants take the first quiz as the pre-test, play the game, evaluate the application by System Usability Scale (SUS), and the last quiz as the post-test. Table 1 displays a summary of the participant's demographics.

Table 1: Main study participant demographics

| Characteristics | Total |
|---|---|
| Sample size | 20 |
| Gender | |
| Male | 16 |
| Female | 04 |
| Age (years) | |
| Age (18-25) | 06 |
| Age (26-35) | 14 |
| Experience using mobile device | |
| Mobile phone | 00 |
| Smartphone | 20 |
| Average coding hours per week | |
| 0-9 | 00 |
| 10-19 | 00 |
| 20+ | 20 |

#### 4.2.2 Procedure

A The think-aloud approach was used to collect data, together with a pre- and post-test. Each participant took about 60 minutes to complete the think-aloud research study using an online Zoom meeting. Each participant was first given a description of the think-aloud experimental study's purpose and were requested to sign a consent form. Before the experiment began, each participant was asked (individually) if they understood what GDPR law meant. Furthermore, participants were told that they could inquire about any concerns they had regarding the experiment. The pre- and post-tests were administered on an Apple MacBook Pro computer, and each test ended with the participants receiving their results.

## 5. Results

### 5.1. SUS Study Results

The objective of the SUS study was to assess the mobile game prototype's general usability. As a result, it used the SUS scoring method introduced by [23]. The SUS generates a single number that serves as a composite indicator of a software product's overall usability (in this case, a gamified browser application). The total score contributions from all items were initially added up to get the SUS score for the mobile game prototype. The score contribution of items 1,3,5,7, and 9 is the scale position minus 1. The score contribution of items 2,4,6,8, and 10 is 5 minus the scale position. Finally, the scores are multiplied by 2.5 to get the overall score of the gamified browser-based application usability. The range of SUS scores is 0 to 100. Hence, in order to accomplish this study, the overall user satisfaction with the developed gamified application was measured using the SUS scale. The scores of the SUS study have been summarized in Table 2.

Table 2: The user satisfaction of the gamified browser-based application

| No | Statement | Average score |
|---|---|---|
| 1 | I think that I would like to use this mobile game frequently | 4.75 |
| 2 | I found the mobile game unnecessarily complex | 1.90 |
| 3 | I thought the mobile game was easy to use | 4.55 |
| 4 | I think that I would need the support of a technical person to be able to use this mobile game | 1.85 |
| 5 | I found the various functions in this mobile game were well integrated | 4.75 |
| 6 | I thought there was too much inconsistency in this mobile game | 1.75 |
| 7 | I would imagine that most people would learn to use the mobile game very quickly | 4.45 |
| 8 | I found the mobile game very awkward to use | 1.85 |
| 9 | I felt very confident using the mobile game | 4.85 |
| 10 | I needed to learn a lot of things before I could get going with this mobile game | 1.80 |
| | **Average satisfaction score range (0-100)** | **81.25** |

Total Score = 32.5
SUS Score = 32 x 2.5 = **81.25**

In general, 81.25 percent of the participants reported subjective satisfaction with the gamified application [23]. To compare the mean of the participants' pre-and post-test scores, the research study used a paired-sample t-test [27]. The survey results indicate high satisfaction with the application.

## 5.2. Think Aloud Study Results

Using the findings from the [31] study as a guide, two stages of data analysis for the think-aloud study were carried out. The study first divided the recordings into different categories by applying keywords to each segment. The keywords were derived from the components of the framework for game design that was presented by [22]. As a result, the findings were classified into twelve categories. Secure coding behavior, motivation, perceived threat, perceived severity, perceived susceptibility, and self-efficacy were the main segments used to categorize the recordings. In the findings of the Think aloud experiment, there was a better result from the post-test as compared to the pre-test. Participants who played the mobile game received pre-test scores of 55% and post-test scores of 88%. GDPR principles were significantly more familiar to participants in the post-test by a margin of 33%.

Figure 3 displays the individual participant's scores during their interaction with the gamified application. In the figure, there was no lower score compared from the pre-test to the post-test, hence indicating a higher improvement. In the secure coding behavior section, one of the participants answered wrongly in the pre-test and eventually answered correctly in the post-test. In the think-aloud experiment, it was noted that the developers played the game and gave out various responses based on how they found it beneficial for privacy protection. One response stated, "Playing a game is better than reading books and articles to learn about privacy implementation." Another response pointed out that "I would love to play the game time and again to learn more." These responses are a clear indication that the keywords were all about learning and privacy protection which are the basis of the experiment toward securing end-user data.

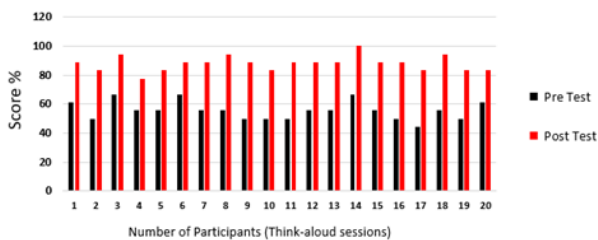

Figure 3: The individual participant's score during their engagement with the mobile game prototype

## 6. Discussion

Research The increasing use and effectiveness of electronic data processing have made data privacy the most crucial issue in the modern day [18]. The protection of personal information is made more difficult by the emergence of future-generation networks [12][13]. In order for developers to incorporate GDPR privacy policies into the software and prevent any future privacy issues, this study experimentally evaluated the game design framework proposed by [22] through a game-based learning application. Twelve areas or categories were given attention, that is Secure Coding Behavior, Motivation, Perceived Threat, Perceived Severity, Perceived Susceptibility, Self-efficacy, lawfulness, fairness and transparency, Purpose Limitation, Data minimization, Integrity and Confidentiality, Accuracy, and Storage Limitation. There was a maximum of twenty sessions and a minimum of thirteen sessions for the twelve areas.

One of the most quoted results in secure coding behavior indicates that the application, in consideration of GDPR principles, assisted developers in preventing privacy threats such as data breaches by limiting access to sensitive data [28]. As for the motivation area, interest in developing privacy-enabled applications that reduce excessive surveillance arose as a result of playing the game [28]. The self-efficacy category noted that the game taught developers GDPR principles. This can be quoted [28] from one of the sessions where a developer said, "This game definitely increased my knowledge about GDPR." The knowledge of GDPR principles enhances the developer's awareness of what steps to take to enable privacy. Under data minimization, developers stated that limited personal data collection was more relevant and necessary and avoided excessive data processing [28]. In the storage limitation category, developers pointed out that the game guided them in determining the data retention period [28].

As a result, the current study makes the straightforward but important point that the game application increases developers' awareness and training in maintaining privacy.

## 7. Conclusion

This study assessed the effectiveness of a game design framework that [22] introduced and validated by [28]. Six GDPR principles were taught to the developers through the design and development of the game, which served as an educational tool.

To address the issue and convey the information in the context of game design, we used a browser-based gamified application. SUS was used as the first step toward assessing the subjective satisfaction of the gamified application. Further, a think-aloud study experiment was conducted with a pre-and post-test to evaluate the game design framework [16]. Right after the pre-test, while playing the game, the participants were asked "exit questions" about the strategies they use to incorporate privacy into their code. Participants' success rate in the pre-test was 55%, while their post-test scores were 88% which translated that the gamified application encouraged learning. Developers' understanding of the GDPR and, consequently, his/her secure coding behavior has grown by a remarkable 33%. As a result, participants were taught how to include privacy-preserving techniques in the software they designed using the mobile game prototype. There were factors that motivate developers to enable privacy which include perceived threat, perceived susceptibility, perceived severity, self-efficacy, lawfulness, fairness and transparency, purpose limitation, data minimization, accuracy, storage limitation, and integrity and confidentiality.